\documentclass[journal]{IEEEtran}

\usepackage{amsfonts,psfrag,epsfig}
\usepackage{amsfonts,epsfig}
\usepackage{color}
\usepackage{amsmath,amssymb,hyperref}
\usepackage[amsmath,thmmarks]{ntheorem}
\usepackage{tabularx}
\usepackage[margin = 1.1in]{geometry}

\hypersetup{colorlinks=true}

\newtheorem{theorem}{Theorem}
\newtheorem{lemma}[theorem]{Lemma}
\newtheorem{corollary}[theorem]{Corollary}

{ \theoremstyle{nonumberplain}\theoremsymbol{\ensuremath{\Box}}
\theorembodyfont{\normalfont}

\theoremsymbol{\ensuremath{\Box}}
\theoremheaderfont{\normalfont\itshape}
\theorembodyfont{\normalfont} \theoremstyle{empty}

}

\newcommand{\cE}{\mathcal{E}}
\newcommand{\cV}{\mathcal{V}}

\newcommand{\beq}{\begin{eqnarray}}
\newcommand{\eeq}{\end{eqnarray}}
\newcommand{\beqn}{\begin{equation}}
\newcommand{\eeqn}{\end{equation}}

\renewcommand{\hat}{\widehat}

\newcommand{\powb}[2]{{\left({#1}\right)}^{{#2}}}

\begin{document}

\title{Efficient Synchronization Stability Metrics for Fault Clearing}

\author{Scott~Backhaus$^1$,
	Russell~Bent$^1$,
       	Daniel~Bienstock$^2$,
	Michael~Chertkov$^1$,
          and~Dvijotham~Krishnamurthy$^3$
\thanks{$^1$Los Alamos National Laboratory (LANL), $^2$Columbia University, $^3$University of Washington.}}%

\maketitle

\IEEEpeerreviewmaketitle

\begin{abstract}
Direct methods can provide rapid screening of the dynamical security of large numbers fault and contingency scenarios by avoiding extensive time simulation. We introduce a computationally-efficient direct method based on optimization that leverages efficient cutting plane techniques. The method considers both unstable equilibrium points and the effects of additional relay tripping on dynamical security\cite{01SH}. Similar to other direct methods, our approach yields conservative results for dynamical security, however, the optimization formulation potentially lends itself to the inclusion of additional constraints to reduce this conservatism.
\end{abstract}

\vspace{-0.5cm}
\section{Introduction}

Exogenous events, e.g. such as faults, generation trips, load trips, and fluctuations of intermittent generation threaten the dynamic stability of power systems, and a large part of power systems operations is devoted to risk assessment, i.e. developing and using online tools  to determine what could happen should an exogenous event occur. The mathematics describing large power systems  is highly complex and non-linear creating significant challenges for developing the computational tools to perform these assessments. The first step in these performing these assessments is dividing the exogenous contingencies and their analysis into subsets of similar structure and behavior---N-1 analysis that checks if there is a feasible steady state following a contingency that changes the structure of the power system  e.g. generator tripping, and transient stability that checks if the post-contingency dynamics lead to the identified steady state. Transient stability must be checked for contingencies that both change power system structure and those that do not, i.e. temporary faults that are automatically cleared. Here, we focus on the dynamics following temporary faults leaving the system structure the same as before the fault. Our emphasis is on developing a computationally efficient method for analyzing the security of these post-fault dynamics.

A large body of previous work has focused on the {\it stability} of post-fault dynamics, i.e. whether the system state leaves the region of attraction surrounding the steady-state minimum in the energy function\cite{85VWC,11Chi,03ZYC,88CWV,95CCC,94CWV}. In contrast, we follow \cite{01SH} where the post-contingency dynamics are {\it secure} if no further tripping of transmission protection devices occurs. Assessing this condition is computationally difficult. Here, we develop a new approach based on convex optimization that results in a conservative assessment.

Similar to the approaches mentioned above, we use time integration to compute the total energy $E_*$ accumulated by the system during the fault-on period. $E_*$ is a combination of kinetic energy $W$ of generators and potential energy $U$ associated transmission line flows. Similar to other stability assessments approaches, we assume  $E_*$ manifests instead as potential energy $U$ with $W=0$. Our approach differs in that it is able to efficiently determine if all of the energetically feasible states ($U\leq E_*$) are also secure with respect to additional protection device operation. If the approach indicates the post-contingency dynamics are secure, then the system is guaranteed stable (under certain assumptions). If the approach indicates that the post-contingency dynamics are insecure, the results are inconclusive.

Often, there are {\it insecure}, low potential energy system states that are the root of the conservatism in all of these approaches. Although these insecure states are energetically feasible ($U\leq E_*$), the post-fault dynamics may never access these states, or if they do, there will be significant kinetic energy associated with these states. An advantage of our approach is that it allows us to find these low energy states, i.e. their configuration of phases. This additional information may allow the development of additional heuristics that remove the much of the conservatism.

The remainder of the paper is organized as follows. In Section \ref{sec:approach} we discuss the overall approach in the context of the existing literature. In Section \ref{sec:formulation} we develop our formal model of dynamic stability and metrics. Section \ref{sec:optimization} develops the optimization algorithm for calculating stability and section \ref{sec:experiments} describes empirical results. We give conclusions and directions for future work in Section \ref{sec:conclusion}.

\vspace{-0.5cm}
\section{Approach}
\vspace{-0.2cm}

\label{sec:approach}

The literature contains a rich history of methods and approaches for assessing the risk associated with faults. The approaches are split into static and dynamic criteria. Static approaches focus on determining if there exists a steady solution to the power flow equations for a given network topology and nodal injections.
This analysis is necessary but not sufficient because it does not test whether the dynamics following a non-steady initial condition lead to the steady solution. For example, a power system following a cleared fault satisfies static criteria as the pre-fault injections are the same as the post fault injections. However, if the fault lasts sufficiently long, the post-fault dynamics may result in large deviations from the steady state causing global instability or, more likely, the tripping of additional protection equipment and significant uncertainty in the resulting dynamical trajectory.

These observations motivate the need to analyze the post-contingency dynamics to determine if they converge to the steady state. A straightforward approach uses extensive time simulations to test all possible contingencies. The number of contingencies is large and there are a large number of degrees of freedom in even moderately large systems creating a computationally difficult task that challenges the use of this method. Instead, we develop efficient algorithms to assess dynamic stability similar to recent work on related direct methods for transient stability \cite{85VWC,95CCC,03ZYC}.

Direct methods \cite{85VWC,95CCC,03ZYC,11Chi} determine stability of a post-fault system based on energy functions that do not require integrating the differential equations that describe the post-fault system. The Potential Energy Boundary Surface (PEBS), \cite{88CWV}, and the Boundary of stability region Controlling Unstable equilibrium point method (BCU) \cite{94CWV} are two well-studied direct methods. The BCU method searches for the controlling Unstable Equilibrium Point (UEP) for a given fault-on trajectory, $x_f(t)$, whose stable manifold (wrt the system dynamics) contains the exit point of the trajectory. PEBS, which uses a lower dimensional approximation of the full system to determine stability boundaries, was developed to circumvent the problem of determining the controlling UEP for a given fault-on trajectory.

Like PEBS, our method is based on the energy formulation and does not require finding the controlling UEP. Unlike PEBS that relies on heuristics, we develop a provably sufficient \footnote{under lossless dynamics and stable voltage assumptions.} stability criterion that is embedded in a convex optimization problem. The convex optimization problem determines the maximum phase difference for each transmission line under the constraint that the potential energy to achieve this phase difference is smaller than the total energy accumulated during the fault. If the phase difference remains below the thresholds of protective equipment, the post-contingency dynamics are secure.

\vspace{-0.2cm}
\section{Dynamic Model}
\label{sec:formulation}

\vspace{-0.1cm}
\subsection{Hamiltonian Dynamics with Damping}
\label{sec:hamil}

The tuple ${\cal G}=({\cal V},{\cal E},\beta)$ defines the vertices, (undirected) edges, and susceptances of a power system, respectively. For lossless lines (resistance is ignored) the system's phase vector $\theta=(\theta_i|i\in{\cal V})$  satisfies the dynamical equations, $\forall i\in{\cal V}$:
\begin{equation}
  M_i\ddot{\theta}_i+\gamma_i\dot{\theta}_i=
 p_i-\sum_{j:\{i,j\}\in {\cal E}}v_i v_j\beta_{ij}\sin(\theta_i-\theta_j),
\label{theta-eq}
\end{equation}
where $p=(p_i|i\in{\cal V})$ is the globally balanced vector of mechanical power inputs and power consumptions. The $M=(M_i|i\in{\cal V})$ are the generators' rotational inertia, and the $\gamma=(\gamma_i|i\in{\cal V})$ represent the generator and load response to local system frequency shifts $\dot{\theta}_i$ via damping and speed droop or via frequency dependent loads. $v_i$ is the voltage at the node $i$, which is assumed to be a tightly controlled constant that potentially varies from node to node.

Eqs.~(\ref{theta-eq}) are restated as a Hamiltonian dynamical system with damping, $\forall i\in{\cal V}$
\begin{eqnarray}
\dot{\theta}_i=\frac{\partial E(\theta,\varpi)}{\partial\varpi_i},\
\dot{\varpi}_i=-\frac{\partial E(\theta,\varpi)}{\partial\theta_i}-\frac{\gamma_i}{M_i}\varpi_i,\nonumber\\ 
E(\theta;\varpi;v;p)=W+U,\ W=\sum_{i\in{\cal V}}\frac{\varpi_i^2}{2M_i},\nonumber\\ 
U=\sum_{\{i,j\}\in{\cal E}}\beta_{ij}v_i v_j(1-\cos(\theta_i-\theta_j)) - \sum_{i\in\cal V} p_i\theta_i,
\label{energy}
\end{eqnarray}
where $\varpi_i=M_i\omega_i$ and $E$ define momentum and total system energy, respectively. The total system energy is composed of  kinetic energy $W$ accumulated in generators' rotation and the system potential energy $U$. When damping is ignored ($\gamma$=0), the energy $E$ is conserved. In the more general case ($\gamma\neq$0), $d E/dt\leq 0$.

\vspace{-0.3cm}
\subsection{Stationary Power Flows and Necessary/Static Synchronization Condition}

For a balanced system at nominal frequency, Eq.~(\ref{theta-eq}) shows that the stationary Power Flow (PF) equations follow from a variation of the potential energy:
$\forall i\in{\cal V},\quad \partial U/\partial \theta_i=0$.
When phase differences over all lines of the system are bounded by $\pi/2$,  $\theta\in \Theta=(\forall_{i,j\in{\cal V}}: \ |\theta_i-\theta_j|\leq \pi/2)$, $U(\theta;v;p)$ is a convex function of $\theta$. When the optimal solution to
$\theta_{\mbox{min}}=\mbox{arg}\min_{\theta \in \Theta} U(\theta;v;p)$
falls within the interior of $\Theta$,  this is the only solution to the PF equations within $\Theta$,  and the dynamical system (\ref{theta-eq}) is stable within the  (possibly infinitesimally small) vicinity of the steady solution. The total energy of the steady solution is $E_{min}= \min_{\theta \in \Theta} U(\theta;v;p)$. If $\theta_{\mbox{min}}$ occurs on the boundary of $\Theta$, then the guarantees of solution existence within $\Theta$ are lost.

\vspace{-0.3cm}
\subsection{Distance Protection Model}
\label{sec:protect}

From a practical power systems perspective, not all $\theta$ in $\Theta$ are feasible with respect to system protection. Specifically, some regions of $\Theta$ will have $\theta_i-\theta_j$ such protection relays may erroneously detect that an additional fault has occurred\cite{01SH}. Following \cite{01SH}, we define a sub-space $\Theta_{relay}$ which lies entirely within $\Theta$. For $\theta \in \Theta_{\mbox{relay}}$, the system will not encounter additional operations of protective devices. We adopt the model of $\Theta_{\mbox{relay}}$ given in the Appendix of \cite{01SH} (simplified here by our assumption of constant $v_i$),
    \begin{equation}
    \Theta_*(\theta^{\max})=\left(\forall \{i,j\}\in{\cal E}:\quad |\theta_i-\theta_j|\leq \theta_{ij}^{\max}\right),
    \label{relay}
    \end{equation}
where $\theta^{\max}=2\arcsin(1/\sqrt{2\beta})$ and $\beta$ (parameter introduced in \cite{01SH}) is a constant describing protective relay's level of security. $\beta=1.2$ corresponds to $\theta_{\mbox{th}}\approx 1.4$ on the rhs of Eq.~(\ref{relay}) and is a typical choice for zone 2 relays.

\subsection{On-Fault Dynamics \& Fault Clearing}
\label{sec:energy}

Prior to the contingency, we assume the system is in a stationary state within $\Theta_{\mbox{relay}}$ and that the stationary state is balanced at the nominal frequency, i.e. at $t=0,\ \forall i\in{\cal V}:\quad\dot{\theta}_i=\ddot{\theta}_i=0$. Prior to the contingency, the total system energy is only the ``stored'' potential energy, $U(\theta^{(\mbox{pre})};v;p)$, where $\theta^{(\mbox{pre})}$ is the pre-contingency phase vector.

In the rest of this manuscript, we simplify the discussion by only considering 3-phase faults so that we can maintain our balanced, positive-sequence representation of the power system. During the fault-on period, we assume that the power system's voltage regulation capabilities are sufficient to maintain constant voltage at the unfaulted nodes while the voltage at faulted node becomes $0$---a condition that eliminates real power flow to this node. During the fault-on period, the dynamics is governed by the equivalent of Eqs.~(\ref{theta-eq}) with the faulted node $k$  and adjacent links removed. The mechanical input powers $p_k$ do not change during the fault-on period, and the imbalance  between the $p_k$ and the network flows causes the generators accelerate and gain kinetic energy $W$. We solve this initial value problem with $\theta(0)=\theta^{\mbox{pre}}$, $\dot{\theta}(0)=0$ evaluating the dynamics over the  fault-on time interval $[0,\tau_f]$, aiming to find $\theta^{(\mbox{post}-)}=\theta(\tau_f^-)$ and $\dot{\theta}^{(\mbox{post}-)}=\dot{\theta}(\tau_f^-)$.

When the fault is cleared, the fault-on network structure reverts to the pre-fault structure. The fault clearing stage is assumed to occur instantaneously so that the frequencies and phases at all the non-faulted nodes $k$ are continuous from $\tau_f^-$ to $\tau_f^+$, i.e.
\begin{eqnarray}
&& \forall k\in{\cal V}\setminus i:\quad \dot{\theta}_k^{(\mbox{post}+)}=\dot{\theta}_k^{(\mbox{post}-)},\label{dot-post}\\
&& \forall k\in{\cal V}\setminus i:\quad \theta_k^{(\mbox{post}+)}=\theta_k^{(\mbox{post}-)}.\label{post}
\end{eqnarray}
At the faulted load node $i$, the frequency is also continuous ($\dot{\theta}_i^{(\mbox{post}+)}=\dot{\theta}_i^{(\mbox{post}-)}$, but the post-fault phase $\theta_i^{(\mbox{post}+)}$  is reconstructed from phases at the neighboring nodes  by resolving
\begin{eqnarray}
p_i=\sum_{j:\{i,j\}\in{\cal E}}v_iv_j\beta_{ij}\sin\left(\theta_i^{(\mbox{post}+)}-\theta_j^{(\mbox{post}-)}\right),
\label{post}
\end{eqnarray}
which assumes that the post-fault power balance at the load node $i$ is established instantaneously.
If the fault has occurred at a generator node, reconstruction of the post-fault phase and frequency follows directly from accounting for the mis-balanced dynamics of the generator during the fault,  thus resulting in
$p_i \tau_f = \frac{1}{2} M_i\powb{\dot{\theta}_i^{(\mbox{post}+)}}{2}$,  where the damping is ignored in comparison with the generator inertia.

These non-stationary, post-fault $\theta$ and $\dot{\theta}$ define the new, post-fault total system energy $E_*$. The pre- and post-fault network structures and the $p_i$ are the same, therefore, the system has the same steady state $\theta_{\mbox{min}}$ corresponding to $E_{min}$. However, the kinetic energy gained by the generators during the fault-on period ensure that $E_*>E_{min}$.

\vspace{-0.3cm}
\subsection{Post-Fault Security as a Convex Optimization Problem}
\label{sec:ConvexOpt}

Our goal is to develop an algorithmically efficient way of determining if the post-fault dynamics initiated by $\theta_*$ and $\dot{\theta}_*=\varpi_{*i}/M_i$ will be secure,  i.e. not resulting asynchronic swings of any of the generators or violation of any of the relay limits (\ref{relay}). We suggest to solve the following new set of convex optimization problems as a conservative method to check for post-fault system security: $\forall \{i,j\}\in{\cal E}$,
\begin{subequations}
\label{bar_theta}
\begin{eqnarray}
\hat{\theta}_{ij} & \doteq & \mbox{arg}\max_{\theta} |\theta_i-\theta_j| \\
\mbox{s.t.} && U(\theta;v;p) \leq E_*(\theta_*;\varpi_*)\\
&& \theta\in\Theta_{\mbox{relay}}.
\end{eqnarray}
\end{subequations}
Optimization (\ref{bar_theta}) is infeasible if $E_*<E_{\min}$.
If $E_*\geq E_{\min}$ and all of the $|{\cal E}|$ optimal solutions lie strictly in the interior of $\Theta_{\mbox{relay}}$, i.e. not on the boundary of $\Theta_{\mbox{relay}}$, then the post-fault dynamics are determined to be secure. $E_{\max}$ marks the largest possible $E_*$ when the solution is (borderline) secure. If $E_*>E_{\max}$ and thus at least one of the $|{\cal E}|$ optimal solutions lies on the boundary of $\Theta_{\mbox{relay}}$, then the respective optimal $\hat{\theta}_{ij}$ corresponds to equality in Eq.~(\ref{relay}) raising the possibility of additional protective relay action.

\begin{figure}[h]
\centering
\includegraphics[width=0.45\textwidth]{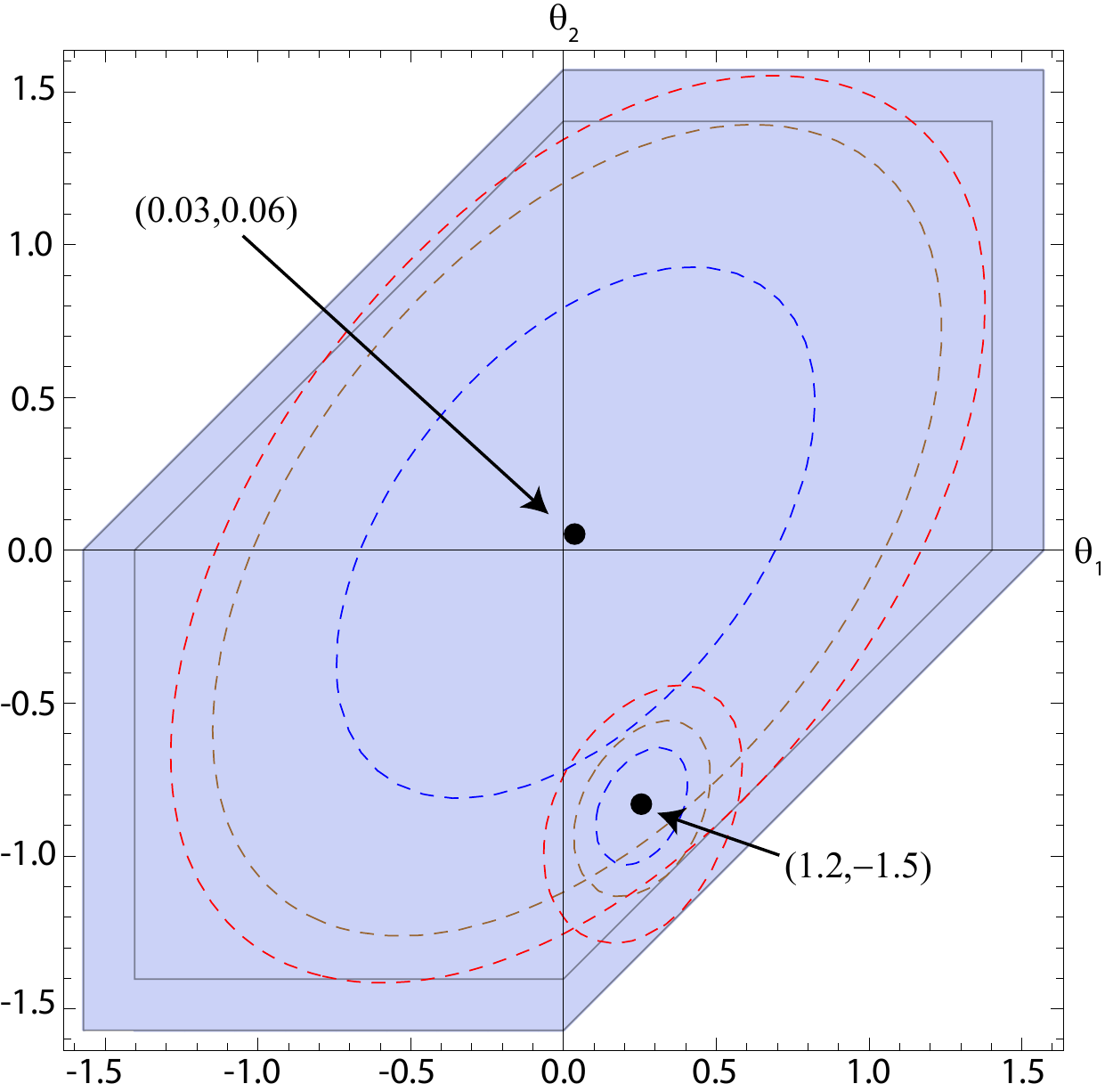}
\caption{ Energy levels versus the two phases of a three node triangle system characterized by potential energy, $U(\theta_1,\theta_2)=(1-\cos(\theta_1))/0.8+(1-\cos(\theta_2))/1.2+
1-\cos(\theta_1-\theta_2)
-p_1\theta_1-p_2\theta_2$.
In the two overlayed cases shown $p_1=0.03, p_2=.06$ and $p_1=1.2, p_2=-1.5$, respectively. In both cases, the bold dot corresponds to the minimum $E_{min}$ of the potential $U$ . The gray-blue colored domain shows $\Theta$ and gray line bounds the sub-domain $\Theta_{\mbox{relay}}$ for $\beta=1.2$. Red and brown dashed lines show iso-lines of the maximum post-fault energy $E_{max}$ that  limit the domains of safe recovery for $\Theta$- and $\Theta_{\mbox{relay}}$-constrained systems respectively. The values for the cases are
$E_{min}\approx 0., E_{max;\Theta}\approx 1.34, E_{max;relay}\approx 1.1$ and $E_{min}\approx -0.7, E_{max;\Theta}\approx -0.67, E_{max;relay}\approx -0.63$ respectively.
\label{fig:three-node}}
\end{figure}

The intuition behind optimization problem (\ref{bar_theta}) is straight forward. The post-fault system with total energy $E_*$ can only access states with potential energy $U$ less than $E_*$, i.e. the domain of all energetically-accessible states is defined by $U(\theta;\tilde{p})\leq E_*(\theta_*;\varpi_*)$. The first constraint in optimization problem (\ref{bar_theta}) restricts the optimal solutions to these energetically-accessible states. This constraint is conservative because not all states with $U(\theta;\tilde{p})\leq E_*(\theta_*;\varpi_*)$ will be visited by the post-fault dynamics. The second constraint in optimization problem (\ref{bar_theta}) further restricts the optimal solution to be within the feasible domain $\Theta_{\mbox{relay}}$ for relay operation and phase synchronization. If optimization problem (\ref{bar_theta}) results in solutions that are only in the {\it interior} of $\Theta_{\mbox{relay}}$, then all energetically-accessible states will not result in additional relay protection operation or loss of synchronization. However, if at least one solution of optimization problem (\ref{bar_theta}) lies on the boundary of $\Theta_{\mbox{relay}}$, then there exists a energetically-accessible state that may result in relay operation or loss of synchronization. Viewed in this manner, optimization problem (\ref{bar_theta}) is a conservative method to assess post-fault security. These conditions are illustrated on a simple three-node case in Fig.~(\ref{fig:three-node}). If the post-fault system has $\theta\notin\Theta$, the system has already encountered one or more additional relay operations, and the solution of Eq.~(\ref{bar_theta}) will not be helpful.

To validate the predictions of the energy-based method described above, we simulate the system dynamics for the post-fault system by solving an initial value problem with the initial state given by $\theta^{(\mbox{post}+)}$ and $\dot{\theta}^{(\mbox{post}+)}$ over the interval $[\tau_f,\tau_{pf}]$, where $\tau_{pf}$ is a simulation time, typically chosen to be 20-30 seconds.

\section{Optimization Algorithm}
\label{sec:optimization}

We consider $\Theta_{\mbox{relay}}=\Theta_*(\theta^{\max})$ and rewrite the optimization in Eq.~(\ref{bar_theta}) as
\begin{subequations}
\label{form2}
\begin{eqnarray}
&&\max \ |\theta_i - \theta_j|,\ \mbox{s.t.} \\
&& \sum_{\{k,h\} \in \cE} \beta_{kh} \psi_{kh}  -  \sum_{k \in \cV} p_k \theta_k  \le  E_*   \\
&&  1 - \cos( \theta_{kh})\le \psi_{kh},\  \forall \{k,h\} \in \cE, \label{psidef} \\
&&  0 \ \le \psi_{kh} \le 1 - \cos(\theta^{\max}_{kh}),\
 \forall \{k,h\} \in \cE \\
&& \theta_1 = 0, \quad  |\theta_{kh}| \le \pi/2, \ \forall \{k,h\} \in \cE, \label{last1}
\end{eqnarray}
\end{subequations}
where $\psi_{kh}$ are newly introduced variables and used that $\sum_{k \in \cV} p_k = 0$ and that the quantities $\theta^{\max}_{kh}$ are at most $\pi/2$. At optimality for problem (\ref{form2}), all constraints
(\ref{psidef}) will hold as equalities, i.e. $1 - \cos( \theta_k - \theta_h )  = \psi_{kh}$ for all lines $kh$. Constraint (\ref{last1}) takes advantage of the fact that $\theta_k$ can be shifted by a common constant and that $\theta^{\max}_{kh} \le \pi/2$ for all lines $\{k,h\}$.

Formulation (\ref{form2}) is nonlinear, but only because of the nonlinear constraint (\ref{psidef}). However (\ref{psidef}) is convex in the problem domain allowing the solution of (\ref{form2}) with a classical cutting-plane algorithm using the lower envelope of the function $1 - \cos(\delta)$. Specifically, the tangent line to $1 - \cos(\delta)$ at $\delta = \delta_0$ is $ \sin(\delta_0)( \delta - \delta_0)+ 1 - \cos(\delta_0)$. Considering the line $\{k,h\}$ and a particular value of $\theta_{kh}$ (i.e. $\tilde \theta_{kh}$ such that $|\tilde \theta_{kh}| \le \pi/2$), we have, by convexity, that $ \sin(\tilde \theta_{kh})(\theta_{kh} - \tilde \theta_{kh})+1 - \cos(\tilde \theta_{kh})\le 1 - \cos(\theta_{kh})$. We can use this observation to approximate formulation (\ref{form2}) by creating a lower bound to (\ref{psidef}) by linearizing about the set of $ \tilde \theta_{kh}$. To create a tighter lower bound on (\ref{psidef}), we employ a series of linearizations at a family of values, $ \tilde \theta_{kh}^{t}, \quad t = 0, \ldots, N_{kh}$, for $\theta_{kh}$.
Thus arriving at the following statement.
\vspace{-0.2cm}
\begin{lemma} \label{stupid} For any line $\{i,j\}$,  $\hat \theta_{ij}$ is at most the value of the Linear Program (LP) derived by replacing in (\ref{form2}) the nonlinear inequality (\ref{psidef}) with the system of linear inequalities, at $t = 0, \ldots, N_{kh}$:
\begin{eqnarray}
\sin(\tilde \theta_{kh}^{(t)})( \theta_{kh} - \tilde \theta_{kh}^{(t)})+1 - \cos(\tilde \theta_{kh}^{(t)}) \le \psi_{kh}. \label{psideflin}
\end{eqnarray}
\end{lemma}
\vspace{-0.2cm}
\noindent {\em Proof.} This follows from the fact that (\ref{psideflin})
is a relaxation of (\ref{psidef}). \QED

In our implementation we solve a sequence of LPs rather than (\ref{form2}). Each problem in the sequence incorporates additional inequalities of the form (\ref{psideflin}). By appropriately choosing the quantities $\tilde \theta^{(t)}_{kh}$ we can attain arbitrarily high accuracy in the approximation. Before formally stating our algorithm, it is important to note the following implication of Lemma \ref{stupid}.
\begin{corollary} \label{verystupid}
Suppose the optimal value is strictly less than $\theta_{ij}^{\max}$.
Then $\hat \theta_{ij} < \theta_{ij}^{\max}$.
\end{corollary}

Corollary \ref{verystupid} is an important ingredient of our algorithm;
in particular, it provides an early termination criterion very useful for improving the algorithm computational efficiency.

Next we describe our algorithm and its reliance on two (small) tolerance parameters $\epsilon$, $\delta$.\\
\underline{\bf Algorithm:}\\
\noindent {\bf Initialization.} An initial LP formulation is
created by choosing a family of positive $\tilde \theta_{kh}^{(t)}$ for each line $\{k,h\}$
\begin{equation}
\tilde \theta_{kh}^{(t)} \ = \ \rho_{kh} \, (1 + \lambda)^{(t-1)},\label{theta_t}
\end{equation}
where $0 < \rho_{kh} = \theta^{\max}_{kh}/10$, $t = 1, 2, \ldots$, and
$\lambda $ is slightly larger than $1.0$. Note that
$\tilde \theta_{kh}^{(1)} = \rho_{kh}$. The highest
value of $t$ used is the largest $t$ such that $\tilde \theta_{kh}^{(t)} \le \theta_{kh}^{\max}$. We set $\tilde \theta_{kh}^{(0)} = 0$. Additionally, we choose negative values $\tilde \theta_{kh}^{(t)}$ which are the precisely the negatives of those in (\ref{theta_t}). We comment on these choices below.

\noindent {\bf Step 1.}  Solve the current LP formulation.

\noindent {\bf Step 2.} {\bf If} the
problem is infeasible, declare it infeasible, and {\bf exit}
the procedure.

\noindent {\bf Step 3.} Otherwise let  $(\psi^*, \theta^*)$ be an optimal solution vector. {\bf If}
$ \theta^*_i - \theta^*_j \ \le \ (1 - \delta) \theta^{\max}_{ij}$,
{\bf then} declare that the optimal value is less than
$\theta^{\max}_{ij}$ and {\bf exit} the procedure.

\noindent {\bf Step 4.} For each line $\{k,h\}$ perform the following task. {\bf If}
$ \psi^*_{kh} + \epsilon < 1 - \cos(\theta_{kh}^*)$,
{\bf then add the cut},
$\sin(\theta_{kh}^*)( \theta_{kh} - \theta_{kh}^*)  +  1 - \cos(\theta_{kh}^*)  \le \psi_{kh}$,
to the formulation, {\bf set}, $ \tilde \theta_{kh}^{N_{kh} + 1}=\theta_{kh}^*$,
and {\bf reset} $N_{kh} \leftarrow N_{kh} + 1$.

\noindent {\bf Step 5}. If no cuts were added in Step 3, {\bf exit.} Else, {\bf Go to 1}.

The following comments are in order\\
\noindent{\bf (a)} As argued above, the formulation
solved at each execution of Step 1 is a relaxation of (\ref{form2}). Thus,
Step 2 is correct, and likewise Step 3 is correct as per Corollary \ref{verystupid}.\\
\noindent{\bf (b)} If condition of Step 4 applies, our
current piecewise-linear approximation to the function $1 - \cos(\theta_{kh})$
is weak at $(\psi^*, \theta^*)$.  Since the left-hand side of the cut inequality in Step 4
takes value $1 - \cos(\theta^*_{kh})$ at $\theta_{kh} = \theta^*_{kh}$, this inequality cuts-off $(\psi^*, \theta^*)$.\\
\noindent{\bf (c)} We can motivate our initialization as follows. By construction, the
chosen $\tilde \theta_{kh}^{(t)}$ are densest near zero because the Taylor series for
the function $1 - \cos(x)$ at $x = 0$, starts with the quadratic term. Thus
a finer mesh is needed for a close approximation near zero.

For completeness, we state the following fact.
\vspace{-0.2cm}
\begin{lemma} \label{basic} The algorithm terminates finitely.\end{lemma}
\vspace{-0.2cm}
\noindent {\em Proof.}  Omitted for brevity.

The description of the Algorithm is in fact a fairly broad
template, and there are three distinct regimes corresponding to three ranges for the value of $E_*$:\\
\noindent {\bf (1)} $E_*$ not near $E_{\min}$ or $E_{\max}$. In such cases
we can use a relatively large value for $\epsilon$, for example $\epsilon = 0.001$. For typical values of $\theta^{\max}_{ij}$ (e.g. not very close to zero) the algorithm will typically terminate in Step 3 after two or three iterations. On the Polish grid\footnote{Case 2746wp, available with MATPOWER \cite{matpowerpaper}; this example has 2746 buses, 520 generators and 3514 lines} our implementation runs in approximately 0.2 CPU seconds on a current workstation and using recent versions of Cplex \cite{CPLEX} or Gurobi \cite{gurobi} to solve the linear programs.\\
\noindent {\bf (2)} $E_*$ smaller than, but close to $E_{\min}$. In this case
problem (\ref{form2}) is infeasible. This case is more challenging because
we must prove infeasibility. As $E_* \rightarrow E_{\min}^{-}$ the feasible
region for (\ref{form2}) shrinks; at $E_* = E_{\min}^{-}$
the feasible region has measure zero. However, even at $E_* = E_{\min}^{-}$
and even when $E_*$ is slightly smaller than $E_{\min}$
our polyhedral relaxation may have \textit{positive} volume if $\epsilon > 0$
too large. However, the following result is key:
\vspace{-0.2cm}
\begin{lemma}. Suppose $E_* < E_{\min}$. Then there exists $\epsilon_0 > 0$
with the property that for any choice of $\epsilon$ with $\epsilon \le \epsilon_0$,  the Algorithm  will terminate in Step 2 proving infeasibility.
\end{lemma}
\vspace{-0.2cm}
\noindent {\em Proof sketch.} Uses continuity of the cosine function. \QED \\
The importance of the Lemma is that we can, indeed use the Algorithm to diagnose
that $E_* < E_{\min}$. This does come at a cost, since the quantity $\epsilon_0$ in the Lemma depends on the relative gap between $E_*$ and $E_{\min}$.
If this gap is very small, we will need to choose $\epsilon$ quite small,
possibly outstripping the numerical capabilities of the linear programming
solver. In our experiments with the Polish grid, such cases require on the order of
one second. This is the total running time, because the infeasibility
only needs to be proved for a single line.\\
\noindent {\bf (3)} $E_*$ close to $E_{\max}$. These cases are similar but not
as stringent as those in case (2). Here, we need to run the cutting
plane with a small value of $\epsilon$ (for example, $\epsilon = 0.0001$).
Ideally we would need to separate those cases where $\hat \theta_{ij}$
is strictly smaller than $\theta_{ij}^{\max}$  from those where the two values are equal. However,
from a practical perspective, the condition $\hat \theta_{ij} \approx \theta_{ij}^{\max}$ is already a sign of system distress, and thus there is less
need for extreme accuracy than in case (2).

\vspace{-0.3cm}
\section{Validation Experiments}
\label{sec:experiments}

We consider an exemplary numerical experiment on the IEEE 118-bus test system \cite{IEEE118} using the base case configuration of load and generation. We apply a fault at a node, and the fault is cleared without the loss of generation, load, or transmission lines. The fault-on dynamics are computed using time simulation of Eqs.~(\ref{theta-eq}) yielding the post-fault state and $E_*$. The dynamic simulations are repeated for different fault durations and different nodes resulting in  different post-fault states and $E_*$. These results are used as inputs for two computations:\\
\noindent (1) \underline{\bf Dynamic Simulation} Time simulation of Eqs.~(\ref{theta-eq}) is used to compute system evolution from the post-fault state. A small amount of damping is introduced at both generators and loads.

\noindent (2) \underline{\bf Efficient Optimization} Eqs.~(\ref{bar_theta}) are solved using $E_*$ and with relay limits of $\pi/8$, i.e. $\Theta_{\mbox{relay}}=\Theta_*(\pi/8)$.
\begin{figure}[h]
\centering
\includegraphics[width=0.45\textwidth]{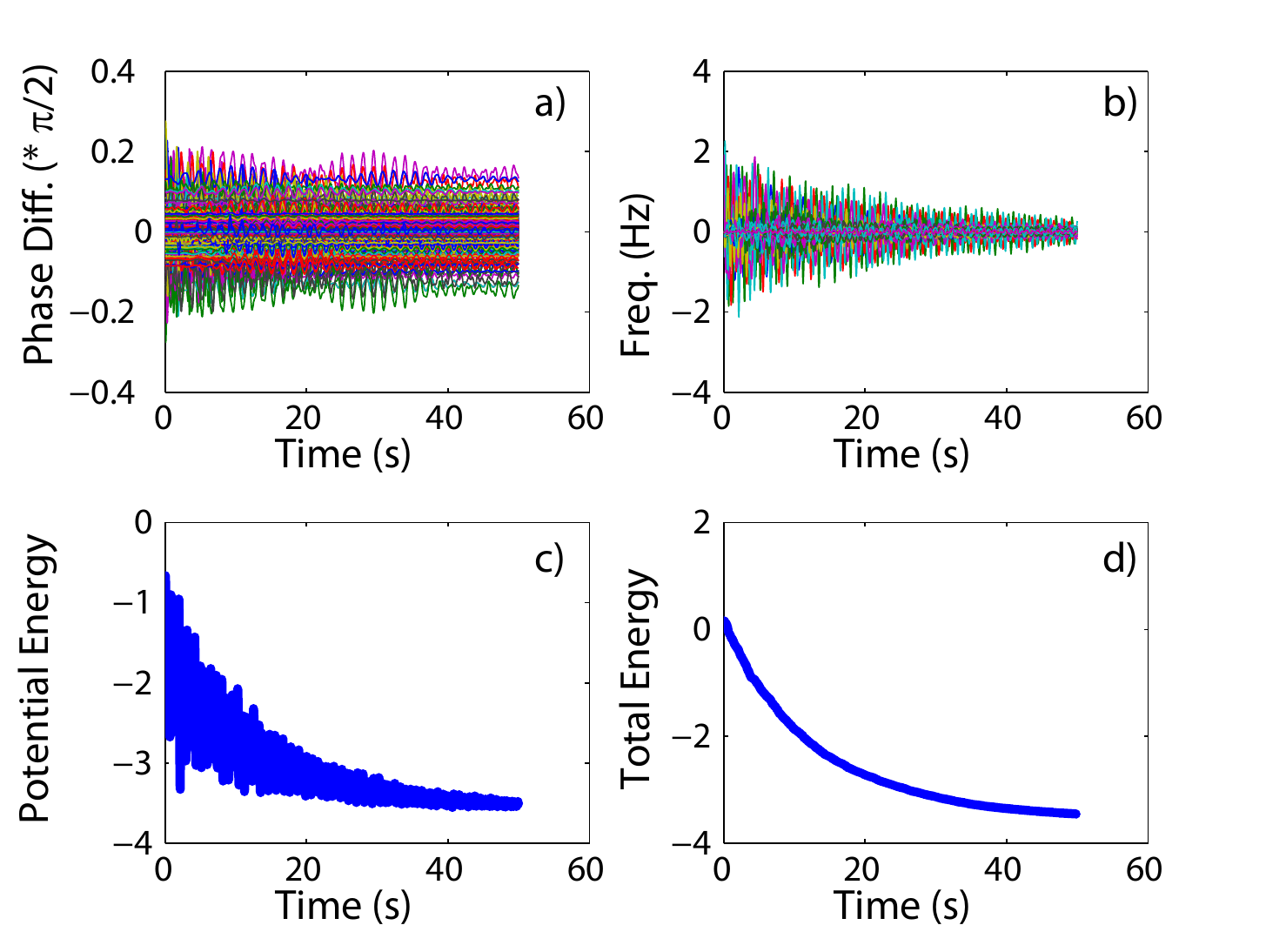}
\caption{The top two sub-figures are the phases and frequencies that are computed from time simulation using the exemplary post-fault state (described in the text) as an initial condition. From the phases and frequencies, the potential energy and total energy are computed and shown in the lower two sub-figures.
\label{fig:post-fault}}
\end{figure}

\begin{figure}[h]
\centering
\includegraphics[width=0.45\textwidth]{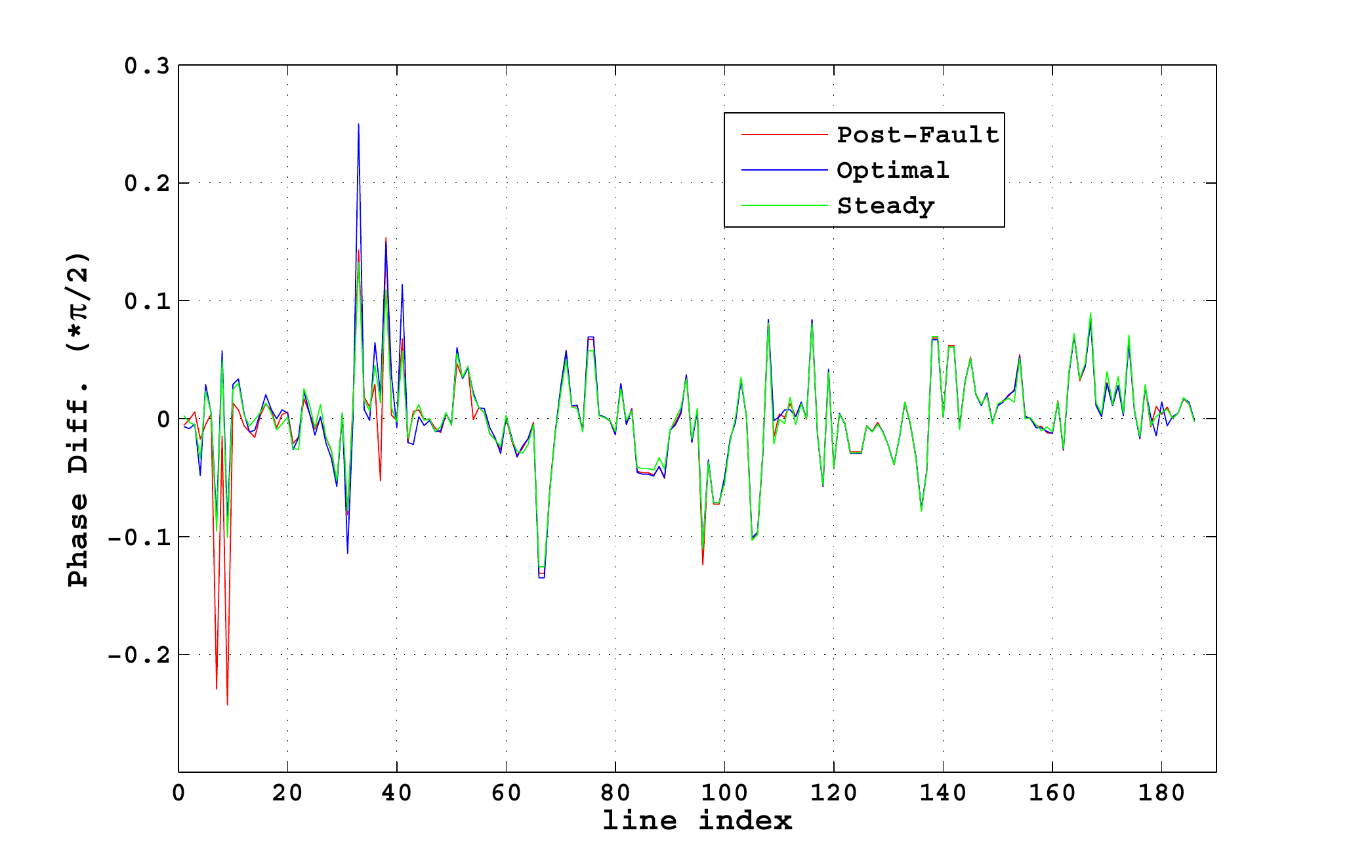}
\caption{Comparison of the post-fault (red), max-energy-optimal (blue), and stationary (green) configurations of phase, where the former two correspondent to initial states for dynamics shown in Fig.~(\ref{fig:post-fault}) and Fig.~(\ref{fig:opt}) respectively.
\label{fig:phases}}
\end{figure}

\begin{figure}[h]
\centering
\includegraphics[width=0.5\textwidth]{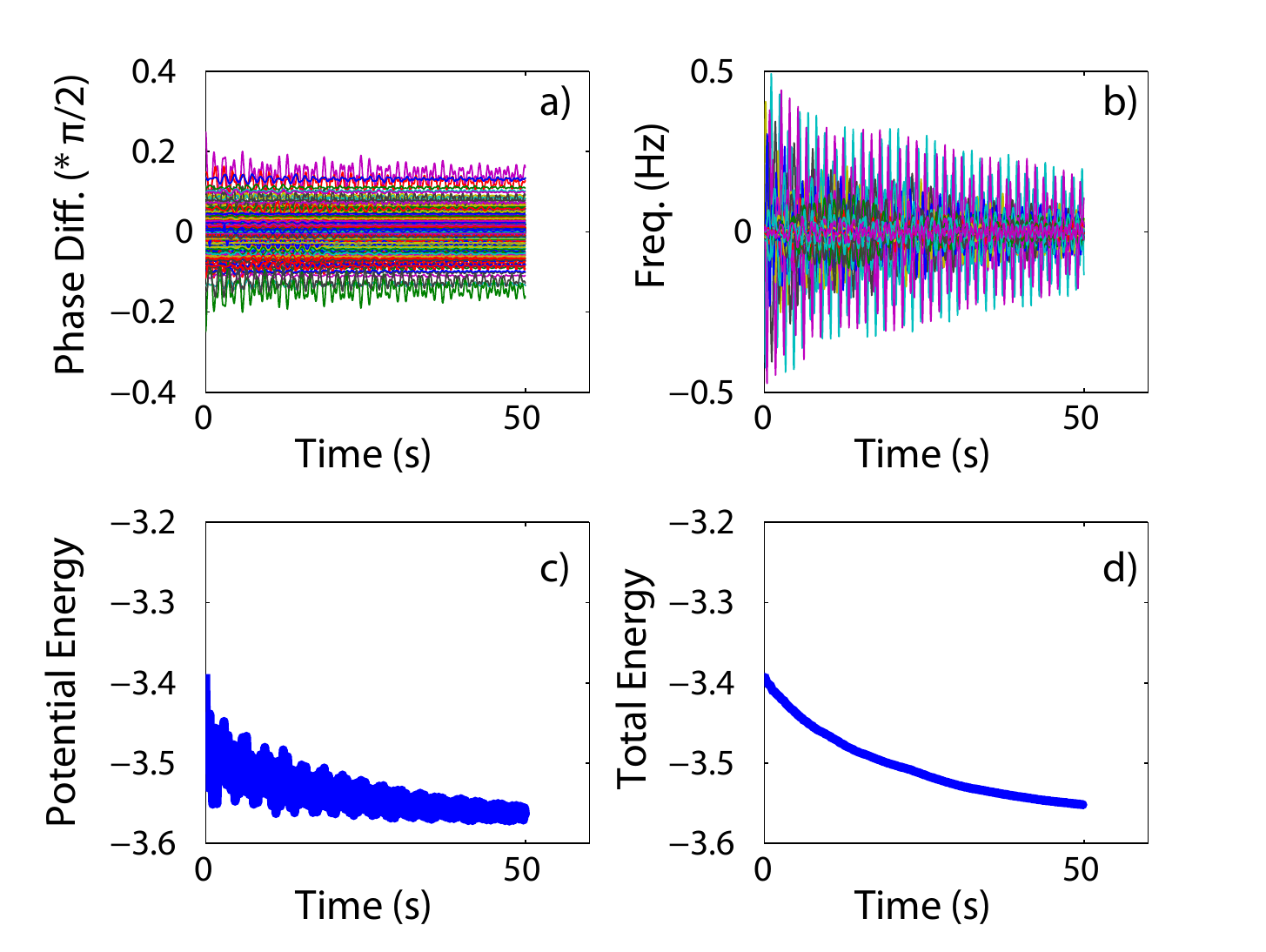}
\caption{All sub-figures are the same as in Fig.~(\ref{fig:post-fault}) except the simulations are
initiated with the configuration of phases optimal for Eqs.~(\ref{bar_theta}) with $\Theta_{\mbox{relay}}(\pi/8)$, $E_*=E_{\max}$, and zero initial $\dot{\theta}$.
\label{fig:opt}}
\end{figure}

For a fault of duration $0.3$ sec at node $\#9$ (the node producing the largest $E_*$ for this duration), $E_*\approx 0$. The post-fault dynamical simulation in Fig.~(\ref{fig:post-fault}) shows the maximum absolute values of the phase difference are close to, but remain below, $\pi/8$ throughout. The small damping introduced into the time simulations in Fig.~(\ref{fig:post-fault}) causes the frequency and phase oscillations to decay. The total energy $E$ also decays, ultimately reaching the steady-state value of $E_{\min} \approx -3.56$. The dynamic simulation shows that the post-fault dynamics are stable and secure. However, the solution of optimization problem (\ref{bar_theta}) for $E^*\approx 0$ and $\Theta_{\mbox{relay}}=\Theta_*(\pi/8)$ yields at least one $\hat{\theta}_{ij}$ on the boundary of $\Theta_{\mbox{relay}}$. This result does not indicate non-secure dynamics. Rather, the result is inconclusive.

These observations have motivated us to experiment with the optimization setting (\ref{bar_theta}) by lowering $E_*$ to find $\bar{E}$, i.e. the largest possible $E_*$ with all $\hat{\theta}_{ij}$ strictly in the interior of $\Theta_{\mbox{relay}}$. We find $\bar{E}\approx -3.4$. Our optimization approach concludes that faults with $-3.56<E_*<-3.4$ result in secure post-fault dynamics (validated by dynamical simulation not shown here). The results of optimization (\ref{bar_theta}) are inconclusive for $E_*>-3.4$, but the dynamical simulations show that, {\it for the faults considered}, the post-fault dynamics are secure revealing the conservative nature of this approach.

To shed some light on this conservatism, we analyze the structure of the solution of Eq.~(\ref{bar_theta}) at $E_*=\bar{E}\approx-3.4$ (see blue curve in Fig.~(\ref{fig:phases})). Indeed, the phase difference across one line is equal to $\pi/8$. For comparison, the steady-state solution ($E=E_{min}=-3.56$) is shown in green and the initial condition for the post-fault dynamics (with $E_*\approx 0$) in Fig.~(\ref{fig:post-fault}) is shown in red. The three solutions are rather similar to each other with significant differences occurring on a few transmission lines. In particular, the $E_*=-3.4$ (blue) and the $E_*=0$ (red) phase differences are qualitatively similar while their difference in total energy is quite large suggesting that the $E_*=0$ state contains a large amount of kinetic energy. This is borne out in Fig.~(\ref{fig:post-fault})b with several of the nodes having significant frequency deviations at $t=0$. This observation suggests that to reach the phase difference configuration of $E_*=-3.4$ immediately or shortly after the fault clearing would also require the state to have a significant amount of kinetic energy. Inclusion of an estimate of this kinetic energy into the total energy could potentially remove a significant amount of conservatism in our direct method.

Next, we investigate the dynamics of the $E_*=\bar{E}=-3.4$ state. Starting from a state of $W=0$, i.e. $\dot{\theta}(0)=0$, Eqs.~(\ref{bar_theta}) are used to compute the phase and frequency dynamics shown in Fig.~(\ref{fig:opt}). Close inspection of the results reveal that one phase difference starts at $\pi/8$ (i.e. 0.25 * $\pi/2$), but the magnitude of all phase differences are quickly contained to less than 0.20 * $\pi/2$. The dynamics appear to quickly exit the region of phase space around this initial condition suggesting that the phase space surrounding this insecure phase configuration is difficult to access in practice.

\vspace{-0.3cm}
\section{Conclusion}
\label{sec:conclusion}

By avoiding the computational burden of extensive simulations, direct methods provide the possibility of rapidly screening the dynamical security of large number of contingency scenarios enabling more frequent assessments of power system security. However, the direct methods themselves must also be computationally efficient. We have formulated a direct method as a computationally efficient optimization problem by utilizing cutting plane methods that can result in early termination of the computation. Or method enables a scalable implementation that can solve the Polish grid case in less than a second. Our method can be used to quickly determine if the post-fault total energy of a particular fault is below a threshold and hence is secure. Alternatively, it can be used to find the lowest energy perturbation and respective dangerous perturbation of the phase vector that result in insecure dynamics which can then be compared against many post-fault total energies.

There are many directions for future work, but following two are perhaps the most important:

\noindent$\bullet$ Reduction of conservatism by inclusion of estimates of kinetic energy in post-fault states or by restriction of the state space of the post-fault dynamics.

\noindent$\bullet$ Inclusion of both phase and voltage dynamics.  In the spirit of our computationally-efficient optimization approach is a robust formulation that accounts for dynamic stability with respect to any spatial configuration of voltage with in a set of nodal-specific confidence intervals.

\bibliographystyle{IEEEtran}
\bibliography{abstract}

\end{document}